\documentclass[9pt,twocolumn,twoside]{optica}
\setboolean{shortarticle}{true}
\setboolean{minireview}{false}

\usepackage[euler]{textgreek}
\usepackage{gensymb}
\usepackage{amsmath}

\let\Lambda\varLambda

\title{High efficiency second harmonic generation of blue light on thin film lithium niobate}

\author[1,*]{Taewon Park}
\author[2]{Hubert S. Stokowski}
\author[2]{Vahid Ansari}
\author[1]{Timothy P. McKenna}
\author[2]{Alexander Y. Hwang}
\author[2]{M. M. Fejer}
\author[2,*]{Amir H. Safavi-Naeini}

\affil[1]{Department of Electrical Engineering and Ginzton Laboratory, Stanford University, Stanford, California 94305, USA}
\affil[2]{Department of Applied Physics and Ginzton Laboratory, Stanford University, Stanford, California 94305, USA}
\affil[*]{Corresponding author: twpark@stanford.edu, safavi@stanford.edu}

\begin{abstract}
We demonstrate second harmonic generation of blue light on an integrated thin-film lithium niobate waveguide and observe a conversion efficiency of $\eta_0= 33000\%/\text{W-cm}^2$, significantly exceeding previous demonstrations. 
\end{abstract}

\setboolean{displaycopyright}{true}

\begin{document}

\maketitle

The strength of interactions between photons in an $\chi^{(2)}$ nonlinear optical waveguide increases at shorter wavelengths. These larger interactions enable coherent spectral translation and light generation at  lower power, over a broader bandwidth, and in a smaller device -- all of which open the door to new technologies spanning fields from atomic clocks~\cite{ludlow2015optical} to optogenetics \cite{segev2016patterned}. Stronger interactions also promise new regimes of quantum optics with rich physics that is only beginning to be considered~\cite{yanagimoto2021towards}.

Here we report highly efficient ($33000\%/\text{W-cm}^2$) second harmonic generation (SHG) of blue light (456.3 nm) on a periodically poled thin film lithium niobate (PP-TFLN) waveguide. To achieve this increase in efficiency, by more than an order of magnitude over previous demonstrations~\cite{wang2018ultrahigh}, we move to shorter wavelengths and reduce the effective mode area. We are able to convert more than 9\% of  1.62 mW laser light in the waveguide to blue light, and verify this by both calibrated measurements of the emitted blue light, and direct observation of 9\% pump power depletion. 

\begin{figure}[htp!]
\centering
\includegraphics{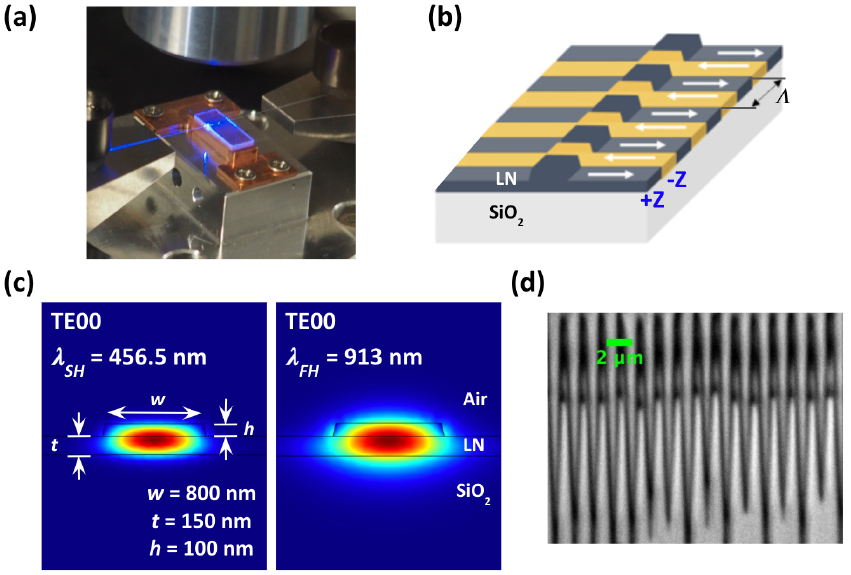}
\caption{(a) On-chip second harmonic generation of blue light captured by visible camera at on-chip pump power $\sim$1 mW. (b) Illustration of the PP-TFLN waveguide. (c) Mode profile for the fundamental transverse electric (TE) mode at $\lambda_\text{FH}$=913 nm and $\lambda_\text{SH}$=456.5 nm. (d) Second-harmonic microscope image of the inverted crystal domain.
}
\label{fig1}
\end{figure}

The design of our device is shown in Figs. \ref{fig1}(b-d). A thin film of X-cut lithium niobate is etched down to form a ridge waveguide. The waveguide width is 800 nm, and we etch 100 nm from 250 nm thick LN film to leave a 150 nm thick slab layer. To achieve efficient conversion, we require quasi-phase matching (QPM) between the fundamental (FH) and the second harmonic (SH) mode. The most efficient conversion occurs via interaction between the two fundamental transverse electric (TE) modes at both wavelengths, and necessitates a first-order poling period of $\Lambda=\frac{\lambda_\text{FH}}{\Delta n_{\mathrm{eff}}}$=1.487 \textmu m.

We obtain this short period ($\sim 1.4~$\textmu m) poling with close to 50\% duty cycle and good uniformity. We start by patterning electrodes on the unprocessed on the MgO-doped X-cut thin-film lithium niobate-on-insulator (LNOI) material (250 nm film of LN bonded to a 2\textmu m layer of silicon dioxide on top of an LN handle wafer by NANOLN). We then use a short pulse with long tail \cite{zhao2020poling} to induce crystal domain inversion. The short portion (<0.3 ms) of the pulse that is above the coercive field strength allows us to avoid domain merging while the long tail (>20 ms) gives enough time for the crystal domain stabilization to result in decent uniformity throughout the 4.2 mm long device verified by SHG microscopy as shown in Fig. \ref{fig1}(d). Finally, we fabricate the waveguides in a similar way to our previous work \cite{mckenna2021ultra}. We pattern resist  using electron beam lithography (JEOL 6300-FS, 100-kV) and transfer the pattern to LN using argon ion milling to obtain the targeted geometry. 

\begin{figure}[htp!]
\centering
\includegraphics{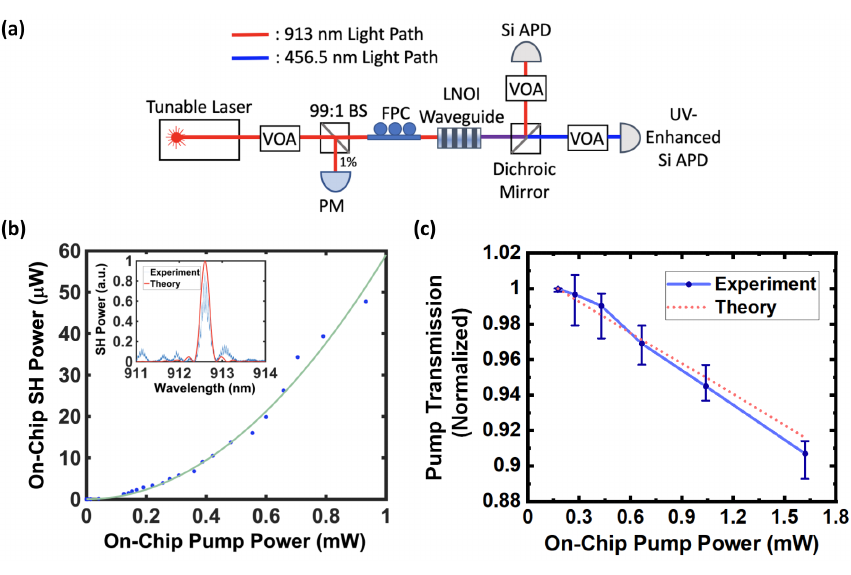}
\caption{(a) Experimental setup for characterizing on-chip second harmonic generation of blue light. VOA (Variable Optical Attenuator), PM (Power Meter), FPC (Fiber Polarization Controller), BS (Beamsplitter). (b) On-chip SH power vs.  pump power, quadratic fit (inset: SHG spectral response, normalized). (d) Normalized pump transmission demonstrating depletion.
}
\label{fig2}
\end{figure}

We characterize second harmonic generation in the device using the experimental setup illustrated in Fig. \ref{fig2}(a). For the input path, we used 780HP fiber where 1\% of the laser light (Velocity TLB-6718, 890-940 nm) goes into the power meter (Newport 918D-UV-OD3R) for pump power calibration. After passing through the fiber polarization controller (FPC), the input pump light is coupled into the chip through 780HP lensed fiber. The light exiting from the chip is collected using 780HP lensed fiber, and then outcoupled to free space. To demultiplex the two wavelengths (i.e., FH and SH), we use shortpass dichroic mirror (Thorlabs DMSP650). The final powers in the FH and SH paths were measured using avalanche photodiodes (Thorlabs APD410A and Thorlabs APD440A2, respectively).

We calculate the theoretical normalized conversion efficiency ($38000\%/\text{W-cm}^2$,  with $d_{33}=25~\text{pm/V}$) using a finite element mode solver (COMSOL). The predicted efficiency is slightly larger than the normalized conversion efficiency obtained from the quadratic fit of the measured data ($33000\%/\text{W-cm}^2$) in Fig. \ref{fig2}(b). This discrepancy in normalized conversion efficiency may be due to waveguide inhomogeneity causing the phase matching condition to vary along the propagation length. There is evidence for this in the SHG spectrum of our device as shown in the inset of Fig. \ref{fig2}(b). The measured response shows a greater spread and reduced peak efficiency as compared to theory.  Integrating this response leads us to believe that with reduced inhomogeneity the peak efficiency could reach $37000\%/\text{W-cm}^2$.  

To estimate the SHG efficiency, we must infer the on-chip SH power from measurements of the out-coupled light. This requires us to understand the fiber-to-chip coupling efficiency at the relevant wavelengths. We characterize the 780HP lensed fiber-to-chip coupling efficiency of the SH light (456.3 nm) with a 532 nm laser (Thorlabs DJ532-10). We use an auxhilary 460HP lensed fiber to characterize the 780HP lensed fiber to chip coupling efficiency at 532 nm ($\sim$11\%), and find that it is slightly higher  than the FH ($\sim$10\%) coupling efficiency. Calibration of the on-chip SH power based on this estimation results in the normalized conversion efficiency of $\eta_0=33000\%/\text{W-cm}^2$ from the quadratic fit of the measured data as shown in Fig. \ref{fig2}(b).

We determine that the propagation loss of our waveguides is below 1dB/cm at the pump wavelength by analyzing fringes on the transmitted light due to back-reflection at the facets. Further and more precise investigation of losses will require fabrication of resonators on this platform~\cite{mckenna2021ultra}. There also does not seem to be any power-dependent loss at the fundamental wavelength, except at wavelengths corresponding to significant SHG, as described below.

To independently verify the normalized conversion efficiency, we measured the transmission of the FH light as a function of power. For this measurement, we used two identical power meters (Newport 918D-UV-OD3R) to monitor the input and the output power of the FH light. The input power meter was connected to the 1\% tap and the output power meter was directly connected to the output 780HP lensed fiber. To remove the effect of the intrinsic nonlinearity of the power meter, we took transmission data at nearby wavelengths (i.e., 910 nm and 915 nm) away from the phase matching wavelength and also with the device out of the loop at different powers. Transmission measurement at the FH wavelength resulted in a $\sim$9.3\% power depletion at input on-chip pump power of 1.62 mW (Fig. \ref{fig2}(c)). This corresponds to the conversion efficiency of $\eta_0=33000\%/\text{W-cm}^2$ which agrees with the calibrated measurement to within our uncertainties. Although the SH light was not filtered out in the output path, the power meter responsivity was $\sim$4 times larger at FH wavelength compared to SH wavelength and the insertion loss of the SH light was $\sim$2 times larger than the FH light. Therefore, the effect of SH light on FH light transmission measurement falls into the range of the inherent uncertainty of the power meter ($\pm$1\%).

In conclusion, we have demonstrated efficient blue light generation on an integrated waveguide. Our work opens up opportunities for a wide variety of applications ranging from quantum optics to bio-sciences.

\medskip

\noindent\textbf{Funding.} Defense Advanced Research Projects Agency (RA-18-02-YFA-ES-578, LUMOS); Nippon Telegraph and Telephone (NTT Research 146395); U.S. Department of Energy (DE-SC0019174); Air Force Office of Scientific Research (MURI No. FA9550-17-1-0002).

\medskip


\bibliography{memorandum}

\begin{thebibliography}{1}
\newcommand{\enquote}[1]{``#1''}

\bibitem{ludlow2015optical}
A.~D. Ludlow, M.~M. Boyd, J.~Ye, E.~Peik, and P.~O. Schmidt, \enquote{Optical
  atomic clocks,} {\protect\JournalTitle{Reviews of Modern Physics}}
  \textbf{87}, 637 (2015).

\bibitem{segev2016patterned}
E.~Segev, J.~Reimer, L.~C. Moreaux, T.~M. Fowler, D.~Chi, W.~D. Sacher, M.~Lo,
  K.~Deisseroth, A.~S. Tolias, A.~Faraon \emph{et~al.}, \enquote{Patterned
  photostimulation via visible-wavelength photonic probes for deep brain
  optogenetics,} {\protect\JournalTitle{Neurophotonics}} \textbf{4}, 011002
  (2016).

\bibitem{yanagimoto2021towards}
R.~Yanagimoto, E.~Ng, T.~Onodera, and H.~Mabuchi, \enquote{Towards an
  engineering framework for ultrafast quantum nonlinear optics,} in
  \emph{Ultrafast Phenomena and Nanophotonics XXV,}  vol. 11684 (International
  Society for Optics and Photonics, 2021), p. 116841D.

\bibitem{wang2018ultrahigh}
C.~Wang, C.~Langrock, A.~Marandi, M.~Jankowski, M.~Zhang, B.~Desiatov, M.~M.
  Fejer, and M.~Lon{\v{c}}ar, \enquote{Ultrahigh-efficiency wavelength
  conversion in nanophotonic periodically poled lithium niobate waveguides,}
  {\protect\JournalTitle{Optica}} \textbf{5}, 1438--1441 (2018).

\bibitem{zhao2020poling}
J.~Zhao, M.~R{\"u}sing, M.~Roeper, L.~M. Eng, and S.~Mookherjea,
  \enquote{Poling thin-film x-cut lithium niobate for quasi-phase matching with
  sub-micrometer periodicity,} {\protect\JournalTitle{Journal of Applied
  Physics}} \textbf{127}, 193104 (2020).

\bibitem{mckenna2021ultra}
T.~P. McKenna, H.~S. Stokowski, V.~Ansari, J.~Mishra, M.~Jankowski, C.~J.
  Sarabalis, J.~F. Herrmann, C.~Langrock, M.~M. Fejer, and A.~H. Safavi-Naeini,
  \enquote{Ultra-low-power second-order nonlinear optics on a chip,}
  {\protect\JournalTitle{arXiv preprint arXiv:2102.05617}}  (2021).

\end{thebibliography}

\bibliographyfullrefs{memorandum}

\end{document}